\documentclass[aps,prd,twocolumn,10pt,nofootinbib,superscriptaddress,floatfix]{revtex4-2}

\usepackage{amsmath,amssymb,bm}
\usepackage{graphicx}
\usepackage{booktabs}
\usepackage[colorlinks=true,citecolor=blue,linkcolor=blue,urlcolor=blue]{hyperref}

\newcommand{\trise}{\tau}
\newcommand{\tauacc}{\tau_{\rm acc}}
\newcommand{\taueng}{\tau_{\rm eng}}
\newcommand{\thetaJ}{\theta_j}
\newcommand{\thetaV}{\theta_v}
\newcommand{\Dkernel}{\mathcal{D}}
\newcommand{\rout}{r_{\rm out}}
\newcommand{\rin}{r_{\rm in}}
\newcommand{\kzero}{k_0}
\newcommand{\Eglobal}{E_{\rm global}}
\newcommand{\Ecoh}{E_{\rm coh}}

\begin{document}

\title{Electromagnetic responses driven by gravitational-wave memory in magnetar-flare outflows}

\author{Wen-Biao Han}
\email{wbhan@shao.ac.cn}
\affiliation{State Key Laboratory of Radio Astronomy and Technology, Shanghai Astronomical Observatory, CAS, 80 Nandan Road, Shanghai 200030, China}
\affiliation{School of Fundamental Physics and Mathematical Sciences, Hangzhou Institute for Advanced Study, University of Chinese Academy of Sciences, Hangzhou 310024, China}
\affiliation{School of Astronomy and Space Science, University of Chinese Academy of Sciences, Beijing 100049, China}
\date{August 8, 2026}

\begin{abstract}
An asymmetric relativistic outflow from a magnetar giant flare produces a
step-like gravitational-wave (GW) memory.  In the magnetized pair plasma
surrounding the star, such GW memory drives transverse electromagnetic (EM)
X-mode/fast-magnetosonic responses at the corresponding frequencies.  We use
this GW-memory spectrum to drive a global wave equation for a cold
electron--positron pair plasma.
With radiative boundary conditions excluding incoming waves, the
driven response forms an outgoing EM mode that reaches the outer boundary of
the source region.  For central-engine energy-release timescales of $10$, $1$,
and $0.15\,\mu\mathrm{s}$, 90\% of the integrated response energy lies below
23.3 kHz, 233 kHz, and 1.51 MHz, respectively.  The principal response
occupies the 10 kHz--1 MHz frequency band; a substantial extension into the
1--3 MHz frequency band also appears when a submicrosecond asymmetric-energy
component is present.  The electromagnetic
response can escape from the source region and potentially propagate in the
interstellar medium.
\end{abstract}

\maketitle

\section{Introduction}
\label{sec:introduction}

Electromagnetic counterparts have become a central part of interpreting
gravitational-wave sources.  GW170817 provided the first GW detection of a
binary neutron-star inspiral and was associated with GRB 170817A
\cite{Abbott2017GW170817}.  Broadband follow-up identified the optical
transient AT~2017gfo and subsequent X-ray and radio emission
\cite{Abbott2017Multimessenger}.  The optical and infrared evolution was
consistent with a kilonova powered by radioactive ejecta and supplied direct
information about the mass and composition of merger outflows
\cite{Tanaka2017}.  Combining the GW luminosity distance with the host-galaxy
redshift also enabled the first standard-siren measurement of the Hubble
constant \cite{Abbott2017Hubble}.  In this conventional counterpart picture,
the GW and EM signals originate from the same astrophysical event, but the EM
radiation is produced by the source matter---through a jet, radioactive ejecta,
or shocks---rather than being driven by the passage of the GW itself.

A GW crossing a magnetic field can also drive an electromagnetic (EM)
disturbance.  The vacuum
limit is the Gertsenshtein effect \cite{Gertsenshtein1962}.  In a plasma the
response may first appear as a coupled matter--field mode.  Linear and
nonlinear treatments have found excitation of Alfv\'en, slow, and fast
magnetosonic branches, including coherent fast-mode growth when the branch is
close to the GW light cone
\cite{BrodinMarklund1999,Servin2000,BrodinDunsby2000,Papadopoulos2001,MoortgatKuijpers2003,Moortgat2004,ServinBrodin2003,Duez2005}.
Relativistic kinetic calculations and numerical strong-field studies retain
the same qualitative coupling while exposing damping and closure dependence
\cite{Isliker2006,Forsberg2010}.  Related work on neutron-star
magnetospheres has considered resonant graviton--photon conversion
\cite{McDonald2024}.

Magnetars store magnetic energies large enough to power recurrent bursts and
rare giant flares \cite{Duncan1992,Thompson1995,Gill2010}.  The 27 December
2004 flare from SGR 1806--20 released an isotropic-equivalent energy of order
$10^{46}\,\mathrm{erg}$ in its initial spike
\cite{Palmer2005,Hurley2005}.  Its resolved radio afterglow expanded,
decelerated, and moved on the sky, providing direct evidence for energetic and
asymmetric ejecta
\cite{Gaensler2005,Taylor2005,Gelfand2005,Granot2006}.  Relativistic,
magnetically dominated, pair-loaded outflows are also natural outcomes of
magnetospheric eruption models
\cite{ThompsonLyutikovKulkarni2002,Lyutikov2006}.  These observations and
models motivate treating the flare not only as an oscillatory stellar source
but also as a rapid, anisotropic redistribution of energy.

Unbound anisotropic motion carries energy and momentum to large radii, changes
the asymptotic gravitational field, and thereby generates linear GW memory
\cite{BraginskyThorne1987,Favata2010}.  The
relativistic-particle result exhibits angular structure and anti-beaming
\cite{SegalisOri2001}; calculations for gamma-ray-burst jets show explicitly
that acceleration or deceleration of a relativistic jet produces a permanent
GW memory \cite{Sago2004,BirnholtzPiran2013,Akiba2013}.  Calculations of
gravitational radiation from accelerating relativistic jets further show that
the observed memory duration depends on both the acceleration of individual
outflow elements and the central-engine activity
\cite{LeiderschneiderPiran2021}.  The asymmetric radio ejecta of SGR
1806--20 therefore provide a concrete physical channel: acceleration and
escape of a net anisotropic relativistic outflow generate a step-like GW
memory.  The observations establish the outflow asymmetry but do not resolve
how rapidly the asymmetric energy was released.  In the model, a shorter
release time extends the GW-memory spectrum to higher frequencies.  Searches for oscillatory GW
transients from magnetar bursts remain important \cite{Corsi2011,Abbott2024}.

The central question of this work is whether the GW memory generated by an
asymmetric magnetar-flare outflow can drive an electromagnetic response that
forms an outgoing wave beyond the source region.  This question has two parts.
The source problem is to determine the finite-frequency GW-memory spectrum
when the outflow acceleration, central-engine energy-release profile, angular
delays, and polarization cancellation are treated separately.  The
propagation problem is to determine whether that spectrum excites only a local
plasma disturbance or a transverse X-mode/fast-magnetosonic response that
continues outward through the magnetosphere and pair outflow.

We address the two parts in sequence.  We first integrate the complex GW
memory of a relativistic top-hat jet, represent the acceleration of individual
outflow elements and central-engine energy release with separate temporal
profiles, and scan the Lorentz factor, acceleration time, opening angle,
viewing angle, and energy-release time.  We then use the resulting GW spectrum
as the distributed driver of a
global non-WKB wave equation for a cold electron--positron pair plasma, with
radiative boundary conditions that exclude incoming waves.  The calculation
spans 0.1 kHz--30 MHz, while the physical interpretation focuses on
10 kHz--3 MHz.  It identifies the central-engine energy-release time as the
main control on the high-frequency extent of the response spectrum and shows that the driven
response forms an outgoing EM mode at the source-region boundary.  The
calculation therefore separates suppression already present in the GW-memory
spectrum from filtering during plasma propagation and establishes the source
conditions required for an escaping 10 kHz--1 MHz response.  The 1--3 MHz
extension requires a submicrosecond asymmetric-energy component and is not
assumed to be universal.

The paper is organized as follows.  Section~\ref{sec:source} constructs the
GW-memory source and magnetar background.  Section~\ref{sec:global} formulates
the local coupling and global wave-transfer problem.  Section~\ref{sec:results}
presents the GW-memory spectrum and outgoing-mode formation.  Section~\ref{sec:discussion}
examines the physical interpretation, model limitations, and possible
observational implications.  Section~\ref{sec:conclusions} summarizes the conclusions.

\section{GW memory and magnetar background}
\label{sec:source}

\subsection{GW memory due to a relativistic top-hat jet}

We represent the asymmetric relativistic outflow by one lobe of a top-hat jet
with half-opening angle $\thetaJ$.  Here ``top-hat'' has its standard geometric
meaning: the energy per unit solid angle and the terminal Lorentz factor are
constant inside the cone and vanish outside it
\cite{Sago2004,BirnholtzPiran2013,Akiba2013}.  Retaining one lobe isolates the
net anisotropic component that produces memory.  It is an effective model for
the observed outflow asymmetry, not an assertion that every magnetar flare
launches an isolated jet without a counterflow.  A symmetric counterjet or a
structured angular profile can be incorporated by adding the corresponding
complex-memory contribution for its orientation and relative energy.

The jet axis is separated from the line of sight by $\thetaV$.  An infinitesimal
angular element of the jet is seen at polar angle $\alpha$ and sky azimuth
$\phi$.  Its complex permanent
memory,
$d{\cal H}=dh_+ + i\,dh_\times$, is
\begin{equation}
d{\cal H}=\frac{2G\epsilon E_{\rm kin}}{c^4r}
\frac{d\Omega}{\Omega_j}
\frac{\beta^2\sin^2\alpha}{1-\beta\cos\alpha}
e^{2i\phi},
\quad \Omega_j=2\pi(1-\cos\thetaJ),
\label{eq:element-memory}
\end{equation}
where $\beta=(1-\Gamma^{-2})^{1/2}$.  The effective asymmetric energy is
$\epsilon E_{\rm kin}$ and is distributed uniformly in solid angle.  The
angular factor is the standard transverse-traceless memory of a relativistic
particle \cite{SegalisOri2001,Sago2004,BirnholtzPiran2013}.  The phase
$e^{2i\phi}$ makes polarization cancellation explicit: a perfectly
axisymmetric jet viewed exactly on axis has zero net complex memory even
though its individual outflow elements are nonzero.

Two physical timescales enter the memory waveform.  The first, $\tauacc$, is
the lab-frame time over which an individual outflow element accelerates.
Retarded arrival times compress it to
\begin{equation}
\tau_\alpha=\tauacc(1-\beta\cos\alpha).
\label{eq:compressed-time}
\end{equation}
For $\alpha\ll1$ and $\Gamma\gg1$,
$\tau_\alpha\simeq\tauacc(\Gamma^{-2}+\alpha^2)/2$.  The second is the
central-engine energy-release timescale $\taueng$, the duration over which the
engine releases the effective asymmetric outflow energy.  Material launched
at different engine times carries that delay into the observed memory, so
$\taueng$ does not receive the same $\Gamma^{-2}$ factor.  This separation
implements the result that jet-memory duration is controlled by both
acceleration and the central-engine energy-release profile
\cite{LeiderschneiderPiran2021}.

For a smooth step, the normalized Fourier transform of its derivative is
\begin{equation}
\Dkernel(f,\tau)=\frac{\pi^2f\tau}{\sinh(\pi^2f\tau)},
\qquad \Dkernel(f,0)=1.
\label{eq:derivative-kernel}
\end{equation}
Representing central-engine energy release and the acceleration of individual
outflow elements by separate smooth temporal profiles gives, for $f>0$,
\begin{equation}
\widetilde h_{\rm M}(f,r)=
-\frac{i}{2\pi f}\Dkernel(f,\taueng)
\int_{\Omega_j}d{\cal H}\,
\Dkernel(f,\tau_\alpha).
\label{eq:jet-memory-spectrum}
\end{equation}
The GW memory is not a monochromatic signal.  In the time domain it approaches
the permanent offset ${\cal H}=\int d{\cal H}$ after the outflow has been
established.  In the frequency domain the ideal step contains a zero-frequency
distribution and a $1/f$ nonzero-frequency component.  The two temporal
profiles in Eq.~\eqref{eq:jet-memory-spectrum} multiply that component by
the corresponding Fourier-domain factors: they leave
$2\pi i f\widetilde h_{\rm M}$ constant at low
frequency and suppress it once a Fourier period becomes shorter than either
physical timescale.  We omit the zero-frequency distribution when calculating
the propagating response; no unique ``memory frequency'' is assigned.

We define the normalized frequency-resolved response energy by
\begin{equation}
{\cal R}_E(f)=
\left|\frac{2\pi if\widetilde h_{\rm M}}{{\cal H}}\right|^2.
\label{eq:normalized-response}
\end{equation}
For comparisons across source models, $f_{1/2}$ denotes the first frequency
at which ${\cal R}_E=1/2$.  It is an operational marker of appreciable
spectral suppression, not a physical cutoff.
The linear GW--fast-mode source is proportional to
$f|\widetilde h_{\rm M}|$ and therefore has a low-frequency plateau.  A
single $10\,\mu\mathrm{s}$ smooth step is recovered when that timescale is
the dominant timescale, for which ${\cal R}_E=1/2$ at $f\tau=0.15111$; it is no
longer assumed to be a universal flare rise time.

\subsection{Angular integration and source-model verification}
\label{sec:angular-integration}

The angular quadrature is uniform in jet azimuth and Gauss--Legendre in
$\mu_j=\cos\vartheta_j$ over $\cos\thetaJ\leq\mu_j\leq1$.  Each node is
rotated by $\thetaV$ into the observer frame to obtain $\alpha$ and $\phi$;
the quadrature weights are normalized so that $\sum_n w_n=1$.  The numerical
form of
Eq.~\eqref{eq:jet-memory-spectrum} is
\begin{equation}
\begin{aligned}
\widetilde h_{\rm M}
&=-\frac{i\Dkernel(f,\taueng)}{2\pi f}
\sum_n w_n h_{0,n}e^{2i\phi_n} \\
&\quad\times\Dkernel[f,\tauacc(1-\beta\cos\alpha_n)] ,
\end{aligned}
\label{eq:discrete-jet}
\end{equation}
where
$h_{0,n}=(2G\epsilon E_{\rm kin}/c^4r)
\beta^2\sin^2\alpha_n/(1-\beta\cos\alpha_n)$.

The angular calculation must reproduce two physically distinct limits.  As
$\thetaJ\rightarrow0$ it approaches the relativistic point-particle result,
whereas an exactly on-axis axisymmetric jet has vanishing net complex memory
because the polarization phases cancel.  Both limits are recovered
numerically.  For the extreme $\Gamma=50$, $\thetaJ=0.03$,
$\thetaV=0.01$ case, the $24\times48$, $48\times96$, and $72\times144$
quadratures all give ${\cal A}=0.099350377$ and
${\cal R}_E(1\,\mathrm{MHz})=0.973216067$ at the reported precision.  For the
representative $\Gamma=10$, $\thetaJ=0.1$, $\thetaV=0.05$ case, the same grids
give ${\cal A}=0.178494459$ and
${\cal R}_E(1\,\mathrm{MHz})=0.015053543$.  Angular refinement therefore
changes none of the reported source conclusions at the 1-percent level.

\subsection{Magnetosphere and pair outflow}

We adopt a magnetar model with radius $R_*=1.2\times10^6\,\mathrm{cm}$, surface field
$B_*=3.0\times10^{15}\,\mathrm{G}$, and spin period $P_*=1\,\mathrm{s}$.
The prescribed field joins $B\propto r^{-3}$ inside the light cylinder to
$B\propto r^{-1}$ in a radial wind.  The inner density is a
Goldreich--Julian-like pair density with multiplicity $\kappa=10^4$
\cite{Goldreich1969}; the outer wind has total pair rate
$\dot N_\pm=10^{38}\,\mathrm{s}^{-1}$ and bulk Lorentz factor
$\Gamma_w=10$.  The source-domain outer boundary is
$\rout=10^{13}\,\mathrm{cm}$.

Strongly magnetized neutron-star plasmas support X, Alfv\'en, and mixed
longitudinal branches whose dispersion and polarization depend on field,
density, temperature, and streaming frame
\cite{AronsBarnard1986,Gedalin1998,Melrose1999,Rafat2019a,Rafat2019b}.
Magnetar magnetospheres also sustain relativistic pair flows
\cite{Beloborodov2013}, and flare-launched Alfv\'en waves can couple to fast
modes \cite{LiZrakeBeloborodov2019}.  As a minimal constitutive closure we use
a cold symmetric electron--positron pair plasma.  In the local wind frame,
\begin{equation}
\omega_p^2=\frac{4\pi n'_{\rm total}e^2}{m_e},\qquad
\omega_B=\frac{eB'}{m_ec},
\label{eq:plasma-frequencies}
\end{equation}
and
\begin{equation}
S(\omega')=1-\frac{\omega_p^2}
{(\omega'+i\nu')^2-\omega_B^2},
\qquad k'^2c^2=\omega'^2S.
\label{eq:dispersion}
\end{equation}
In a symmetric pair plasma the electron and positron contributions to the
gyrotropic off-diagonal response cancel, whereas their diagonal transverse
response adds.  The resulting branch is electromagnetic in polarization and
continuously approaches the vacuum light wave as $S\rightarrow1$.  In the
strong-field limit $\omega_B^2\gg\omega'^2$, Eq.~\eqref{eq:dispersion} gives
$S\simeq1+\omega_p^2/\omega_B^2$; density and magnetization therefore enter
through their competition rather than through a simple unmagnetized plasma
cutoff.  We use the notation X-mode/fast-magnetosonic response to emphasize
this continuous transverse, field-carrying branch across the magnetosphere
and wind descriptions.  It denotes one modeled response, not the sum of two
independent signals.

The radial Lorentz transform is applied before solving for the outward,
attenuating lab-frame root $k_X(r,f)$.  The baseline collision fraction is
zero.  This closure isolates whether smooth dispersion and nonuniformity alone
destroy coherence; thermal, kinetic, and flare-induced damping are deferred.

\section{GW driving and global wave transfer}
\label{sec:global}

\subsection{Local source normalization}

For GW propagation perpendicular to the background magnetic field
($\theta_B=\pi/2$), the coupling coefficient is
\begin{equation}
C_{gF}(f,\theta_B)=\frac{2\pi f\sin\theta_B}{2c},
\qquad \theta_B=\frac{\pi}{2}.
\label{eq:coupling}
\end{equation}
It has the frequency and angular dependence of uniform-MHD fast-mode driving
\cite{Papadopoulos2001,Moortgat2004}.  With the rescaled magnetic Fourier
amplitude $U=r\widetilde B$, define
\begin{equation}
q(r,f)=r C_{gF}(f)B_0(r)\widetilde h_{\rm M}(f,r).
\label{eq:source-term}
\end{equation}
This form exposes the origin of the spectral plateau.  At frequencies below
both source turnovers, the memory spectrum scales as
$|\widetilde h_{\rm M}|\propto f^{-1}$, while the linear coupling contributes
one power of $f$.  Their product is therefore approximately independent of
frequency.  The background magnetic field supplies the dimensional conversion
scale and makes the inner magnetosphere dominate the GW--EM coupling.  The
factor of $r$ removes the leading spherical dilution from the propagated
magnetic amplitude.  The complex phase of $q$ is retained, so contributions generated
at different radii can interfere in the global solution rather than being
added as positive local energies.

For comparison, the local transport equation is
\begin{align}
\frac{dU}{dr}&=q-
\left(\frac{\alpha}{2}+i\Delta k\right)U,
\label{eq:local-amplitude}\\
\Delta k&=\mathrm{Re}\,k_X-\frac{2\pi f}{c},
\qquad
\alpha=2\,\mathrm{Im}\,k_X.
\label{eq:local-dispersion}
\end{align}
Setting $\Delta k=\alpha=0$ gives the coherent local approximation; retaining
both gives the local approximation including dispersion and damping.  These
local approximations provide useful comparisons with the global calculation,
but they retain only the outward-propagating component and cannot describe
reflection.

\subsection{Non-WKB global boundary-value problem}

The rescaled global field $\Psi=r\widetilde B$ satisfies
\begin{equation}
\frac{d^2\Psi}{dr^2}+k_X^2(r,f)\Psi
=2i\kzero q(r,f)e^{i\kzero(r-\rin)},
\qquad \kzero=\frac{2\pi f}{c}.
\label{eq:global-wave}
\end{equation}
Unlike Eq.~\eqref{eq:local-amplitude}, Eq.~\eqref{eq:global-wave} retains both
radial wave directions.  A varying $k_X$ can therefore generate reflection,
and large local wavelengths do not invalidate the propagation method.  The
constitutive relation $k_X(r,f)$ remains local; the calculation is spatially
global but not a kinetic plasma simulation.

No wave is injected into the computational domain.  We impose radiative
boundary conditions excluding incoming waves.  With the spatial convention
$e^{ikr}$ for an outward wave, these conditions are
\begin{align}
\Psi'(\rin)+ik_X(\rin)\Psi(\rin)&=0,
\label{eq:inner-bc}\\
\Psi'(\rout)-ik_X(\rout)\Psi(\rout)&=0.
\label{eq:outer-bc}
\end{align}
The inner condition permits a source-generated inward wave to leave the domain
toward the star but forbids an outward wave from entering through $\rin$; the
outer condition admits only an outward wave.  These two radiation conditions
make the outgoing amplitude an output of the distributed source rather than
an imposed initial value.

\subsection{Exact shell propagation and radiative-boundary solution}
\label{sec:shell-propagation}

The radial domain is divided into shells on which $k_X$ and $q$ are held at
their midpoint values.  Within shell $j$, define the augmented state
$\bm y=(\Psi,\Psi',\phi)^T$, where
$\phi=e^{i\kzero(r-\rin)}$.  The inhomogeneous second-order equation becomes
the homogeneous first-order system
\begin{equation}
\frac{d\bm y}{dr}=\bm M_j\bm y,
\qquad
\bm M_j=
\begin{pmatrix}
0&1&0\\
-k_j^2&0&2i\kzero q_j\\
0&0&i\kzero
\end{pmatrix}.
\label{eq:shell-matrix}
\end{equation}
The exact shell update is
$\bm y_{j+1}=\exp(\bm M_j\Delta r_j)\bm y_j$.  This construction propagates
the vacuum phase analytically inside each shell; the radial grid resolves the
background and source variation rather than sampling every field oscillation.

Let $\bm y_b$ be the outer state of the particular solution initialized with
$(0,0,1)^T$, and let $\bm y_s$ be the outer state of the independent
homogeneous solution initialized with $(1,-ik_{\rm in},0)^T$.  The inner
complex amplitude required by the outer radiation condition is
\begin{equation}
A_{\rm in}=-
\frac{y'_{b}-ik_{\rm out}y_{b}}
{y'_{s}-ik_{\rm out}y_{s}}.
\label{eq:inner-amplitude}
\end{equation}
Propagating $(A_{\rm in},-ik_{\rm in}A_{\rm in},1)^T$ then satisfies both
radiation conditions.  The outgoing complex amplitude at the outer boundary is
$\Psi(\rout)e^{-i\kzero(\rout-\rin)}$.

In uniform vacuum, $k_X=\kzero$ and constant $q$, the numerical solution
reproduces
\begin{equation}
\Psi(\rout)e^{-i\kzero(\rout-\rin)}
=q(\rout-\rin),
\label{eq:coherent-benchmark}
\end{equation}
recovering the coherent first-order result.  A zero source returns zero field.
These two analytic limits validate the normalization and boundary signs.

The one-sided source-boundary spectral energies used below are
\begin{align}
\frac{dE_{\rm EM}}{df\,d\Omega}&=
\frac{2\rout^2c}{4\pi}|\widetilde B(\rout,f)|^2,
\label{eq:em-energy}\\
\frac{dE_{\rm GW}}{df\,d\Omega}&=
\frac{\rout^2c^3(2\pi f)^2}{8\pi G}
|\widetilde h(\rout,f)|^2.
\label{eq:gw-energy}
\end{align}
The EM energy inherits the normalization of Eq.~\eqref{eq:coupling}; ratios
between the global solution and the local approximations are more robust than
its absolute value.

\subsection{Numerical verification and convergence strategy}
\label{sec:numerical-verification}

The source and propagation calculations are verified independently before
they are combined.  The jet calculation is tested against the narrow-jet point
limit, exact on-axis axisymmetric cancellation, the $f\rightarrow0$
permanent-memory limit, monotonic energy-release suppression, and angular
quadrature refinement.  The global calculation is tested against the
zero-source and uniform-vacuum analytic limits, vanishing transverse coupling,
and the two radiation conditions.

The radial convergence study uses 300, 600, 1200, and 2400 radial
shells at ten representative frequencies spanning the full frequency range.
Because every piecewise-constant shell is
advanced by its exact matrix exponential, this refinement tests the radial
representation of the prescribed magnetosphere, wind transition, and source
profile.  It does not impose the unnecessary requirement that the shortest
electromagnetic wavelength be resolved cycle by cycle.  We additionally
monitor radiation-boundary residuals at every selected frequency.

\section{GW-memory spectrum and formation of an outgoing electromagnetic mode}
\label{sec:results}

The calculation separates two questions that need not have the same answer.
The source model determines how much GW-memory power is available at each
nonzero frequency, while the global plasma calculation determines whether the
driven response remains an outward-propagating mode across the source region.
We first identify the source timescales and geometric factors that populate
the kilohertz-to-megahertz spectrum, and then test the transfer of those
components through the prescribed magnetosphere and pair outflow.

\subsection{Relativistic timescales and jet-geometry dependence of the
GW-memory spectrum}

Here and below, ``acceleration-only'' denotes an impulsive central-engine
energy release ($\taueng=0$), leaving $\tauacc$ as the only finite source
timescale.
The source-parameter scan covers 0.1 kHz--30 MHz with 250 unique frequencies,
$\Gamma=5,10,20,$ and 50, $\tauacc=30,100,$ and $300\,\mu\mathrm{s}$,
$\thetaJ=0.03,0.1,$ and $0.3$ rad, and 768 acceleration-only configurations.
The angular quadrature uses $48\times96$ nodes.  For an exactly on-axis
axisymmetric jet, the complex polarization contributions cancel in the
angular integral and the net permanent memory vanishes.  We therefore
interpret the spectral extent only for configurations with nonzero net memory
after angular integration and report both the spectral shape and the amplitude
ratio
\begin{equation}
{\cal A}=\frac{|{\cal H}|}
{(2G\epsilon E_{\rm kin}/c^4r)\beta^2},
\label{eq:amplitude-ratio}
\end{equation}
where the denominator is the point-particle memory evaluated at
$\alpha=\pi/2$, which we call the side-on point-particle reference.  This
fixed normalization removes the common energy--distance scale and the leading
$\beta^2$ dependence, so that ${\cal A}$ isolates the effects of jet angular
extent, viewing geometry, and polarization cancellation.

\begin{figure*}
\includegraphics[width=0.98\textwidth]{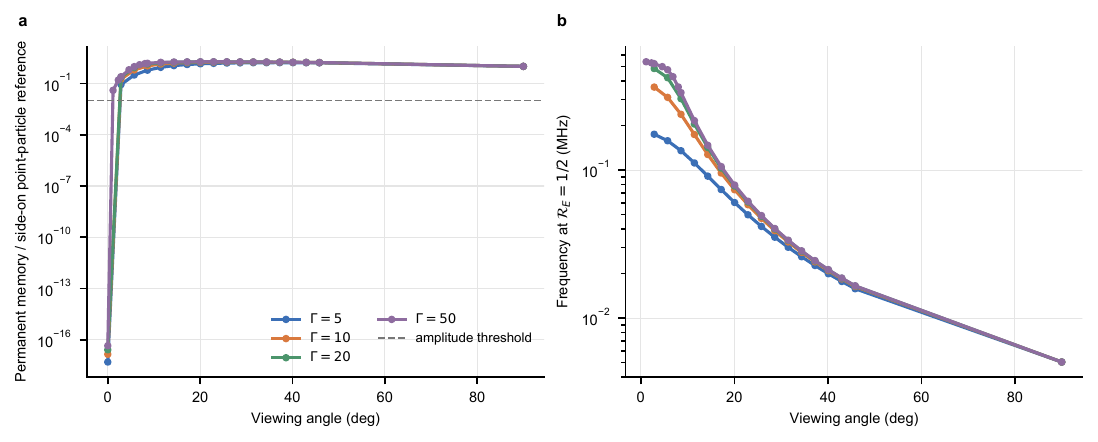}
\caption{\textbf{Viewing-angle dependence of GW memory due to a top-hat jet.}
(a) Permanent complex-memory amplitude relative to the side-on point-particle
reference defined in Eq.~\eqref{eq:amplitude-ratio}, for $\thetaJ=0.1$ rad
and $\tauacc=30\,\mu\mathrm{s}$.  The dashed line marks the amplitude threshold
${\cal A}=0.01$.  At $\thetaV=0$, axisymmetry produces exact on-axis
polarization cancellation.  (b) Frequency at which the acceleration-only
response falls to one-half of its low-frequency value.  Higher Lorentz factors
strengthen relativistic arrival-time compression near the jet axis, but this
effect weakens as the viewing angle increases.}
\label{fig:jet-viewing}
\end{figure*}

For $\tauacc=30\,\mu\mathrm{s}$ and $\thetaJ=0.1$ rad, increasing $\Gamma$
extends the acceleration-only response to higher frequencies near the jet while the exact on-axis
amplitude remains zero (Fig.~\ref{fig:jet-viewing}).  At
$\Gamma=10$, $\thetaV=0.05$ rad, the angularly integrated top-hat model has
${\cal A}=0.178$, and its response falls to one-half of the low-frequency
value at 0.364 MHz.  Across the full scan and requiring ${\cal A}\geq0.1$,
this factor-of-two decline occurs no higher than 0.228, 0.825, 2.469, and
4.989 MHz for $\Gamma=5,10,20,$ and 50, respectively.  These comparison
points describe a continuous decline rather than a cutoff.  The last value has
$\tauacc=30\,\mu\mathrm{s}$, $\thetaJ=0.03$ rad, and
$\thetaV=0.015$ rad, with ${\cal A}=0.220$.  Thus the MHz-scale response is
not created solely by dividing by a vanishing memory amplitude.

\begin{figure*}
\centering
\includegraphics[width=0.45\textwidth]{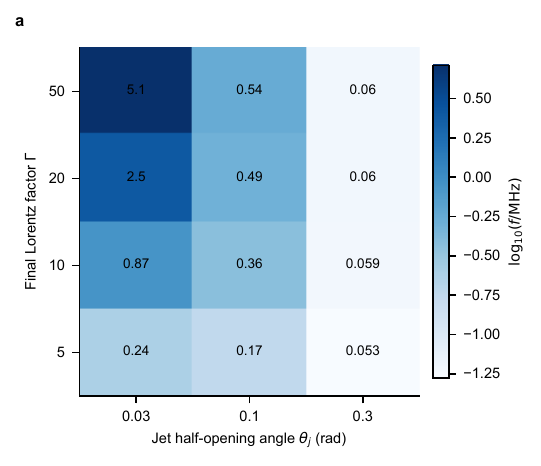}
\caption{\textbf{Dependence of the spectral response on jet parameters.}
For each final Lorentz factor $\Gamma$ and jet half-opening angle $\thetaJ$,
the cells give the first frequency (in MHz) at which the acceleration-only
response falls to one-half of its low-frequency value.  At each grid point,
this frequency is maximized over $\tauacc$ and $\thetaV$ for
${\cal A}\geq0.01$.  Within the scanned grid, this frequency increases toward
smaller $\thetaJ$ and larger $\Gamma$.}
\label{fig:jet-envelope}
\end{figure*}

Jet opening angle limits the gain even before the central-engine energy-release profile is
included
(Fig.~\ref{fig:jet-envelope}).  With the less restrictive ${\cal A}\geq0.01$
threshold, the response remains above one-half of its low-frequency value up to at
most 5.12 MHz for $\thetaJ=0.03$ rad, 0.54 MHz for $\thetaJ=0.1$ rad, and
0.06 MHz for $\thetaJ=0.3$ rad.  A broad top-hat contains many outflow elements with larger
$\alpha$; their longer arrival times and complex polarization factors erase
the point-particle $\Gamma^{-2}$ scaling after angular integration.  Angular
integration therefore strongly limits the high-frequency response of broad
jets, whereas increasing $\Gamma$ extends the response to higher frequencies
only for narrow jets.

The viewing-angle scan separates three geometric regimes.  Exact alignment
maximizes arrival-time compression for individual outflow elements but gives zero net
memory for an axisymmetric top-hat because all polarization phases are
represented equally.  Slightly off-axis configurations retain much of the
compression while breaking that cancellation, producing spectra with the
greatest high-frequency extent at a measurable fraction of the side-on amplitude.  Farther from
the jet axis, the permanent-memory amplitude can remain substantial, but the
characteristic angle $\alpha$ is too large for strong relativistic time
compression.  The physically useful region is therefore a near-edge
compromise between high-frequency extent and amplitude, rather than the formal on-axis
maximum of the point-particle estimate.

\subsection{Central-engine control of spectral suppression}

The acceleration-only spectrum assumes that the entire effective asymmetric
energy is released impulsively.  A finite central-engine energy-release
profile enters through the independent Fourier-domain factor
$\Dkernel(f,\taueng)$ in Eq.~\eqref{eq:jet-memory-spectrum}.  To isolate this
effect, we use the acceleration-only configuration with the largest
$f_{1/2}$ among cases satisfying ${\cal A}\geq0.01$ as the reference case.
Its parameters are $\Gamma=50$, $\tauacc=30\,\mu\mathrm{s}$,
$\thetaJ=0.03$ rad, and $\thetaV=0.01$ rad.

\begin{figure*}[t]
\includegraphics[width=0.98\textwidth]{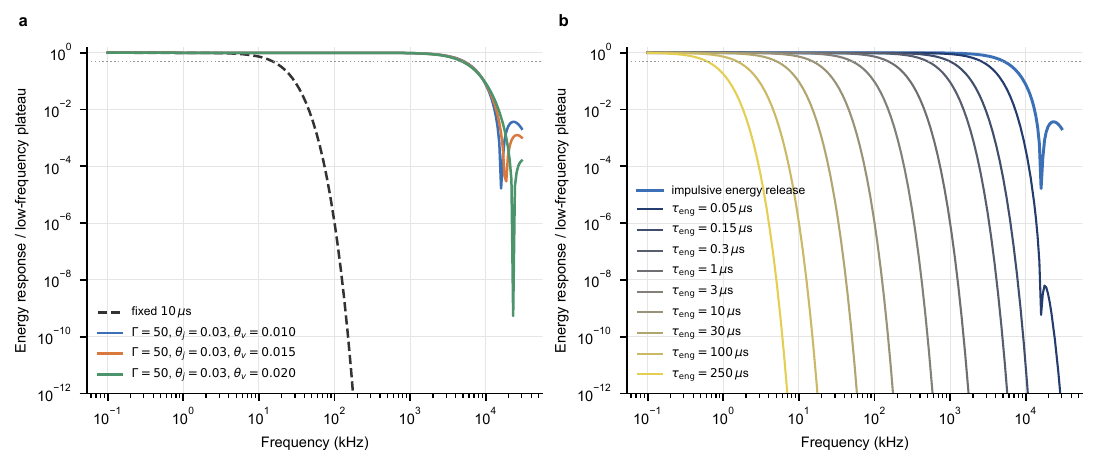}
\caption{\textbf{Relativistic arrival-time compression and the central-engine
energy-release timescale.}
(a) Normalized acceleration-only energy response for three narrow-jet
$\Gamma=50$ configurations, compared with the normalized energy response
$\Dkernel^2(f,10\,\mu\mathrm{s})$ of a single smooth step (dashed;
Eqs.~\eqref{eq:derivative-kernel} and \eqref{eq:normalized-response}).  The
dotted line marks
${\cal R}_E=1/2$.  (b) For the reference configuration with nonzero net memory,
increasing $\taueng$ shifts the spectral decline to lower frequencies.}
\label{fig:jet-engine-timescales}
\end{figure*}

Figure~\ref{fig:jet-engine-timescales}(a) compares three acceleration-only
spectra with the normalized response of a single smooth
$10\,\mu\mathrm{s}$ step.  Panel (b) shows that increasing $\taueng$ shifts
the spectral decline to lower frequencies.  For the reference configuration,
the frequency at which ${\cal R}_E=1/2$ follows
$f\taueng\simeq0.15111$ across $\taueng=0.05$--$250\,\mu\mathrm{s}$.
The central-engine energy-release profile therefore determines where
appreciable spectral suppression begins.

Retaining at least one-half of the low-frequency response at 1 MHz requires
$\taueng\lesssim0.15\,\mu\mathrm{s}$, for which the corresponding light-travel
distance is $c\taueng\lesssim45$ m.  No observation or global magnetar-flare
simulation currently establishes that most of the asymmetric outflow energy is
released on such a short timescale.  The spectra calculated with submicrosecond
$\taueng$ assume that the full asymmetric energy participates in the fast
component.  If the participating fraction is $\eta_{\rm fast}$, the
high-frequency memory amplitude and linear EM-response energy scale as
$\eta_{\rm fast}$ and $\eta_{\rm fast}^2$, respectively.  For
$E_{\rm kin}=10^{46}\,\mathrm{erg}$, $\epsilon=0.1$, and a distance of 10 kpc,
the permanent-memory amplitude is $5.3\times10^{-28}$ for the 5.12-MHz extreme
case.  It is $9.5\times10^{-28}$ for the representative $\Gamma=10$,
$\thetaJ=0.1$, and $\thetaV=0.05$ case.  These are GW-memory amplitudes, not
EM-detector strains or fluxes.

\subsection{Frequency distribution of the response energy}

\begin{figure*}[t]
\includegraphics[width=0.98\textwidth]{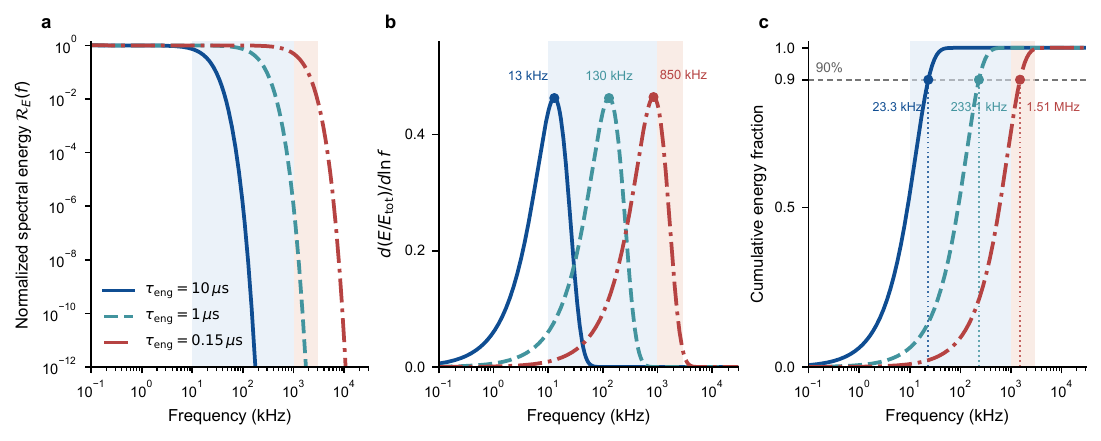}
\caption{\textbf{Frequency distribution of the GW-memory response energy.}
(a) Normalized spectral response energy ${\cal R}_E(f)$ per unit linear
frequency for central-engine energy-release times of 10, 1, and
$0.15\,\mu\mathrm{s}$.  (b) Normalized energy fraction per logarithmic
frequency interval.  Colored markers show broad maxima near 13 kHz, 130 kHz,
and 0.85 MHz.  (c) Cumulative fraction of the integrated response energy;
colored markers and vertical dotted lines show $f_{90}=23.3$ kHz, 233 kHz,
and 1.51 MHz.  The shaded regions mark the 10 kHz--1 MHz and 1--3 MHz
frequency bands.  All panels use the reference configuration with nonzero net
memory: $\Gamma=50$,
$\tauacc=30\,\mu\mathrm{s}$,
$\thetaJ=0.03$ rad, and $\thetaV=0.01$ rad.}
\label{fig:source-energy-spectrum}
\end{figure*}

We next quantify how the response energy is distributed in frequency for
three representative central-engine energy-release times.  Figure~\ref{fig:source-energy-spectrum}(a)
shows ${\cal R}_E(f)$ for $\taueng=10$, $1$, and
$0.15\,\mu\mathrm{s}$.  Each spectrum approaches a constant at low
frequencies and declines at a frequency set by $\taueng$.  Because
${\cal R}_E(f)$ gives the response energy per unit linear frequency, the
low-frequency plateau does not imply a preferred zero-frequency response or
divergent low-frequency energy.

The corresponding normalized energy per logarithmic interval is
\begin{equation*}
\frac{d(E/E_{\rm tot})}{d\ln f}
=\frac{f{\cal R}_E(f)}{\int_0^\infty {\cal R}_E(f')\,df'}.
\end{equation*}
This quantity vanishes as $f\rightarrow0$ and integrates to unity over
$d\ln f$.  In Fig.~\ref{fig:source-energy-spectrum}(b), its broad maxima
occur near 13 kHz, 130 kHz, and 0.85 MHz for the three values of $\taueng$,
respectively.  These maxima show a characteristic spectral scale set by
$\taueng$, but they do not represent a resonance of the GW--EM conversion.

To characterize the cumulative spectrum without relying on a single chosen
level crossing, we define the 90-percent energy frequency $f_{90}$ by
\begin{equation}
\int_0^{f_{90}}{\cal R}_E(f)\,df
=0.9\int_0^{\infty}{\cal R}_E(f)\,df .
\label{eq:f90}
\end{equation}
The analytic limit ${\cal R}_E(0)=1$ supplies the zero-frequency value.  The
upper integral is evaluated to 30 MHz, beyond which the tail is negligible
for the three cases.  Figure~\ref{fig:source-energy-spectrum}(c) shows the
cumulative energy fraction.  For $\taueng=10$, $1$, and
$0.15\,\mu\mathrm{s}$, we obtain $f_{90}=23.3$ kHz, 233 kHz, and 1.51 MHz,
respectively.  The 10 kHz--1 MHz frequency band contains 45.8\%, 94.0\%, and
72.9\% of the corresponding integrated energy.  For the
$0.15\,\mu\mathrm{s}$ case, a further 25.8\% lies between 1 and 3 MHz,
whereas only 0.32\% lies above 3 MHz.  Thus 10 kHz--1 MHz is the principal
frequency band, but 1--3 MHz is a non-negligible extension for a
submicrosecond energy-release timescale.

Table~\ref{tab:engine-summary} summarizes the three representative
central-engine energy-release times and their spectral properties.  For each
$\taueng$, $c\taueng$ gives the corresponding light-travel distance.  The
comparison shows that an appreciable response in the megahertz range requires
submicrosecond asymmetric-energy release, rather than only a large bulk
Lorentz factor.

\begin{table}
\caption{Representative central-engine timescales and corresponding GW-memory
spectral properties for the reference configuration with nonzero net memory.
The last column is the fraction of the integrated response energy in the
10 kHz--1 MHz frequency band.}
\label{tab:engine-summary}
\begin{ruledtabular}
\begin{tabular}{cccc}
$\taueng$ & $c\taueng$ & $f_{90}$
& $E_{10\,\mathrm{kHz}-1\,\mathrm{MHz}}/E_{\rm tot}$ \\
\hline
$10\,\mu\mathrm{s}$ & $3.0\,\mathrm{km}$ & $23.3\,\mathrm{kHz}$ & 45.8\% \\
$1\,\mu\mathrm{s}$ & $300\,\mathrm{m}$ & $233\,\mathrm{kHz}$ & 94.0\% \\
$0.15\,\mu\mathrm{s}$ & $45\,\mathrm{m}$ & $1.51\,\mathrm{MHz}$ & 72.9\% \\
\end{tabular}
\end{ruledtabular}
\end{table}

\subsection{The single-timescale limit and the 0.1--300-kHz reference band}

\begin{figure*}[t]
\includegraphics[width=0.98\textwidth]{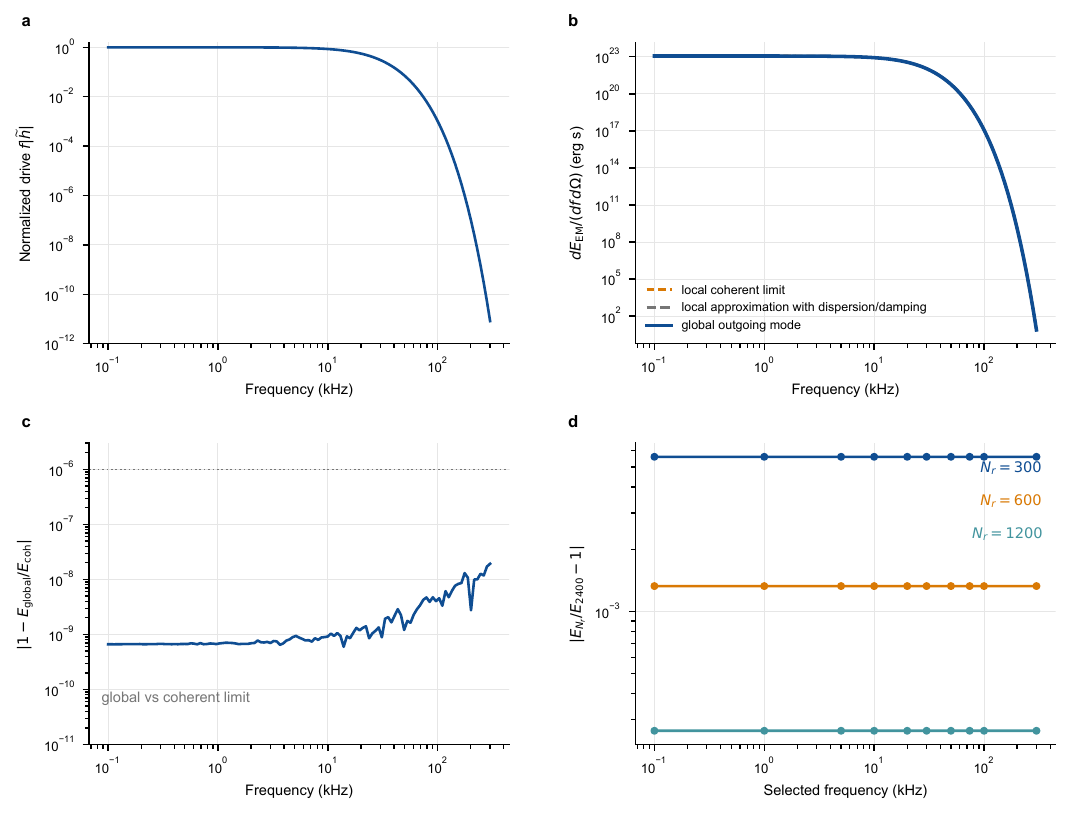}
\caption{\textbf{Global source-region transfer of the GW-memory-driven response.}
(a) Normalized nonzero-frequency drive $f|\widetilde h|$ for a GW memory with a
rise time of $10\,\mu\mathrm{s}$.  The frequency scale of the high-frequency
decline is set by the GW-memory rise time.  (b) Source-boundary EM spectral energy
from the global boundary-value problem, coherent local approximation, and
local approximation including dispersion and damping.  The three curves
overlap at the displayed scale.
(c) Fractional global-energy departure from the coherent limit.
(d) Fractional differences in the outgoing EM spectral energy obtained with
300, 600, and 1200 radial shells relative to the 2400-shell result.}
\label{fig:global-transfer}
\end{figure*}

To connect the GW-memory spectrum calculation with the global propagation
calculation, we first consider a single $10\,\mu\mathrm{s}$ memory sampled at
121 logarithmic frequencies from 0.1 to 300 kHz with 1200 logarithmic radial
shells.  The solution is also evaluated
directly at ten selected frequencies: 0.1, 1, 5, 10, 20, 30, 50, 74, 100, and
300 kHz.  The normalized drive
$f|\widetilde h|$ is flat for $f\trise\ll1$ and falls exponentially for
$f\trise\gtrsim1$ (Fig.~\ref{fig:global-transfer}(a)).  For the analytic
memory spectrum, ${\cal R}_E=1/2$ at $f\trise=0.15111$, corresponding to
15.11 kHz when $\trise=10\,\mu\mathrm{s}$.
The response energy retains 73.0, 31.3, and 9.45 percent of its value at 0.1 kHz
at 10, 20, and 30 kHz, respectively.  It falls to $5.04\times10^{-3}$ at
50 kHz and $1.04\times10^{-6}$ at 100 kHz.  These ratios follow the smooth
memory rise and are not plasma resonances.

\begin{figure*}[t]
\includegraphics[width=0.98\textwidth]{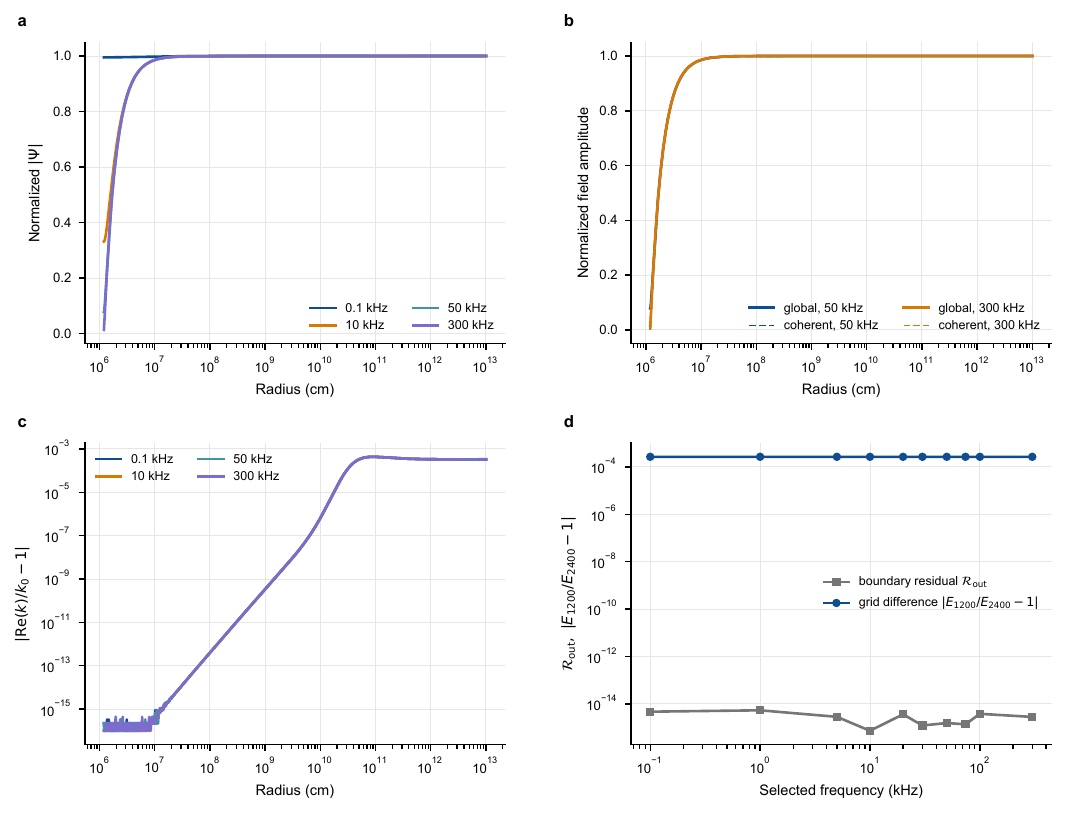}
\caption{\textbf{Radial formation and source-region propagation of the outgoing EM response.}
(a) Radial magnitude of the rescaled global field, $|\Psi(r,f)|$, normalized by
its value at the outer boundary, for four representative frequencies.  The
finite amplitude at the inner boundary is the inward-propagating part of the
source response, which is allowed to leave the computational domain.
(b) Comparison between the global solution and the coherent local approximation
at 50 and 300 kHz.  (c) Radial variation of the X-mode refractive-index departure
from vacuum, $|\mathrm{Re}(k_X)/k_0-1|$.  (d) Numerical accuracy of the global
solution, quantified by the outgoing-boundary residual ${\cal R}_{\rm out}$ and
the relative selected-frequency energy difference between $N_r=1200$ and 2400,
$|E_{1200}/E_{2400}-1|$.}
\label{fig:profiles}
\end{figure*}

\subsection{Negligible source-region suppression of the electromagnetic
response}

This subsection tests whether the GW-memory-driven EM response reaches the
outer boundary $\rout$ as an outward-propagating mode, rather than being
appreciably attenuated or reflected within the source region.  At 0.1 kHz,
the global boundary-value solution gives an outgoing EM spectral energy
$\Eglobal=1.15124\times10^{23}\,\mathrm{erg\,Hz^{-1}\,sr^{-1}}$ in the model
normalization.  Both $\Eglobal(f)$ and $\Ecoh(f)$ denote
$dE_{\rm EM}/(df\,d\Omega)$ evaluated at $\rout$.  The former is obtained from
the global boundary-value solution, whereas the latter is obtained from the
coherent local approximation with $\Delta k=\alpha=0$.  Their ratio measures
how source-region propagation changes the outgoing EM spectral energy relative
to fully coherent outward accumulation.  Across the complete 0.1--300 kHz grid,
\begin{equation}
0.99999998075\leq\frac{\Eglobal}{\Ecoh}
\leq0.99999999940.
\label{eq:transfer-range}
\end{equation}
Equation~\eqref{eq:transfer-range} shows that the global outgoing energy
differs from the coherent result by at most $1.93\times10^{-8}$
(Fig.~\ref{fig:global-transfer}(c)).  The local
approximation including dispersion and damping is similarly close to the
coherent result.  Thus, within the adopted background, cumulative phase
mismatch, damping, and reflection remove no measurable fraction of the
outgoing response.

We next tested whether a plasma barrier appears at higher frequencies by
extending the transfer calculation from 0.1 kHz to 30 MHz.  The calculation
used 161 frequencies, with convergence checks at 1, 3, 10, and 30 MHz.  A
$0.05\,\mu\mathrm{s}$ reference spectrum was used to avoid loss of numerical
precision from the rapidly decreasing high-frequency source amplitude.  In
the linear problem, the global-to-coherent ratio is independent of the overall
source amplitude.  Its largest formal departure from unity is
$1.85\times10^{-6}$, below the maximum 1200-to-2400-shell difference of
$2.66\times10^{-4}$.  The largest radiation-boundary residual is
$4.8\times10^{-15}$.  The extended transfer ratio is therefore consistent with
unity at the numerical resolution of the calculation.

Between 0.1 and 300 kHz, the response energy inherited from the GW-memory
spectrum varies by more than 22 orders of magnitude, whereas the
source-region transfer ratio changes only in the eighth decimal place.  The
high-frequency decline is therefore inherited from the GW-memory spectrum
rather than produced by plasma propagation.  Within the adopted magnetar
background, the response at every computed frequency from 0.1 kHz to 30 MHz
reaches $\rout$ as an outgoing EM mode and propagates beyond the modeled source
region, with no appreciable source-region suppression or additional
plasma-selected cutoff.

\subsection{Radial formation of the outgoing electromagnetic mode}

The magnitude of the rescaled global field $|\Psi|$ grows primarily between the stellar surface
and approximately $10^7\,\mathrm{cm}$ and is nearly constant farther out
(Fig.~\ref{fig:profiles}(a)).  This behavior is shared across the displayed
frequencies after normalization.  At 50 and 300 kHz the global and coherent
radial amplitudes are indistinguishable at line width
(Fig.~\ref{fig:profiles}(b)).  The reason is visible in
Fig.~\ref{fig:profiles}(c): at the selected frequencies the real refractive index
differs from vacuum by at most $6.77\times10^{-9}$.  A smooth feature near the
magnetosphere--wind transition does not accumulate enough phase to suppress
the outgoing wave.

The normalized outer-boundary residual is below
$1.3\times10^{-14}$ in every 0.1--300-kHz convergence run.  Comparing 1200 with 2400
radial shells, the largest selected-frequency energy change is
$2.65\times10^{-4}$; for 600 shells it is $1.33\times10^{-3}$.  The conclusion
of negligible source-region suppression is robust because even the
conservative refinement change is orders of magnitude below an order-unity
spectral filter.

\section{Physical interpretation and model limitations}
\label{sec:discussion}
\label{sec:implications}

\subsection{From a permanent memory to an outgoing wave}

The mechanism is most clearly understood as a sequence rather than as a
conversion between two monochromatic waves.  Asymmetric relativistic ejecta
change the asymptotic metric and leave a viewing-angle-dependent permanent GW
memory.  Because this memory develops over a finite time, its Fourier
transform contains nonzero-frequency components in addition to the
zero-frequency distribution associated with the final displacement.  The
linear plasma equations respond independently to each of these Fourier
components.  Thus ``the same frequency'' means that a component of the
GW-memory spectrum at $f$ drives the X-mode/fast-magnetosonic response at that
same $f$; it does not assign a single oscillation frequency to the memory as a
whole.

This construction is related to, but not identical with, vacuum
Gertsenshtein conversion.  In vacuum the generated field can be described as
an electromagnetic wave accumulating coherently along a prescribed magnetic
path \cite{Gertsenshtein1962}.  Here that magnetic field is embedded in a
nonuniform pair plasma, so the initially driven object is a plasma eigenmode
with both field and matter content \cite{Papadopoulos2001,Moortgat2004}.  The
global calculation follows that mode until its outward field component reaches
the source boundary.  Because the background and coupling are linear and
stationary, the calculation preserves Fourier frequency; it contains no
nonlinear up-conversion.  Any subsequent conversion of the escaping
kilohertz--megahertz response to radio frequencies would be a separate physical
stage.

The GW-memory spectrum calculation and the global propagation calculation play
complementary roles.  The GW-memory spectrum determines how strongly each
frequency drives the plasma response.  The global boundary-value problem
determines whether the driven disturbance is a
local, partly reflected plasma oscillation or an outgoing mode that reaches
the edge of the source region.  In the adopted background the latter occurs:
the outgoing EM response is established mainly inside $10^7\,\mathrm{cm}$ and is
transported to $\rout$ with a global-to-coherent energy ratio indistinguishable
from unity at the accuracy of the calculation.  This transfer cannot restore
high-frequency components already suppressed in the GW-memory spectrum, and it
does not select a narrower preferred band.

The broad maxima in Fig.~\ref{fig:source-energy-spectrum}(b) are set by the
central-engine energy-release timescale rather than by a resonance of the
GW--EM conversion.

The GW-memory spectrum calculation also shows why replacing a nominal flare
time by $\tauacc/(2\Gamma^2)$ is insufficient for a complete outflow.  Arrival-time
compression applies to the acceleration of a relativistic outflow element
close to the line of sight.  Angular integration adds outflow elements with longer delays and
complex polarization factors, while material launched at different engine
times retains the uncompressed timescale $\taueng$.  Consequently, 10--30 kHz is
natural when the effective asymmetric energy is released over several to tens
of microseconds.  An appreciable response at hundreds of kilohertz or at
megahertz frequencies instead requires a submicrosecond component that carries
a non-negligible fraction of the asymmetric energy.  The quantile $f_{90}$
describes this shift without introducing an artificial cutoff at 1 MHz.

This high-frequency extension is therefore a physical requirement to be
tested, not a generic prediction for magnetar giant flares.  A dynamical model
that provides $\taueng$, $\tauacc$, $\Gamma$, $\thetaJ$, and the fast
participating fraction $\eta_{\rm fast}$ can be inserted directly into
Eq.~\eqref{eq:jet-memory-spectrum}.  A response extending to 1 MHz requires
$\taueng\lesssim0.15\,\mu\mathrm{s}$ and appreciable asymmetric energy released
on this timescale.  If $\taueng$ is set by the light-crossing time of a single
causally connected engine region, the corresponding causal scale would be of
order 45 m.  If global flare calculations instead give
$\taueng\gtrsim10\,\mu\mathrm{s}$, relativistic acceleration by itself cannot
prevent the response from declining above the 15-kHz scale.

\subsection{Assumptions and limitations of the source and plasma models}

The top-hat jet is a controlled representation of the net asymmetric outflow,
not a complete magnetar-ejecta solution.  It assumes one effective lobe, a
constant energy per unit solid angle and terminal Lorentz factor inside the
cone, a common acceleration time, and separable smooth temporal profiles for
acceleration and central-engine energy release.  A counterjet, angular structure, a distributed
radial acceleration profile, precession, or time-dependent magnetization can
change both ${\cal H}$ and its spectrum.  The asymmetry parameter $\epsilon$
is likewise an effective projected quantity, and the model does not determine
$\eta_{\rm fast}$ from first-principles flare dynamics.  The parameter scan
therefore maps the kinematic conditions required for a GW-memory spectrum
extending to high frequencies; it is not an event-population or rate
calculation.

Equation~\eqref{eq:global-wave} removes the local-WKB assumption from the
radial propagation problem because the global equation retains both wave
directions, reflection, and exact shell phases.  The constitutive closure is
nevertheless local and deliberately minimal.  The wavenumber is obtained from
a cold, symmetric, one-dimensional electron--positron pair plasma with a
prescribed magnetic field and density.  A hot distribution can modify mode
content and damping \cite{Melrose1999,Rafat2019a}; charge imbalance, oblique
propagation, QED vacuum polarization, nonlinear mode coupling, and a
flare-created pair fireball are omitted.  The scalar source normalization also
does not supply the eigenvector overlap and absolute energy normalization of a
self-consistent kinetic calculation.  The agreement between $\Eglobal$ and
$\Ecoh$ is therefore a statement about source-region transfer within this
closure, not proof of high conversion efficiency in every flare environment.

A more complete calculation should obtain the outflow timescales, angular
structure, and $\eta_{\rm fast}$ from relativistic magnetar-flare dynamics and
project the GW driving onto normalized plasma eigenmodes.  It should then
evolve the thermal pair distribution, field geometry, and flare loading before
matching the source-region solution to an asymptotically free EM wave.
Systematic variations of $B_*$, $\kappa$, $\dot N_\pm$, and $\Gamma_w$ would
identify the backgrounds in which the near-vacuum transfer found here ceases
to hold.  Equations~\eqref{eq:jet-memory-spectrum} and
\eqref{eq:global-wave}, together with the angular and radial convergence tests,
provide benchmarks for those extensions.

\section{Conclusions}
\label{sec:conclusions}
\label{sec:conclusion}

We have followed the complete source-region chain from asymmetric relativistic
ejecta to an outgoing electromagnetic response.  A magnetar giant flare first
produces a step-like GW memory because the unbound outflow changes the
asymptotic gravitational field.  Representing the net asymmetric component as
a relativistic top-hat jet allows the viewing-angle dependence, polarization
cancellation, and arrival-time compression to be integrated explicitly.  The
memory waveform is shaped by two distinct temporal processes: acceleration of
each outflow element and release of asymmetric energy by the central engine.
The Fourier-domain factors associated with their temporal profiles determine
the nonzero-frequency GW spectrum that drives the plasma.

The GW-memory spectrum calculation establishes a hierarchy of spectral scales.
Angular integration limits the point-particle Lorentz compression, and the
uncompressed central-engine energy-release profile sets the high-frequency extent once it
is included.  For $\taueng=10$, $1$, and $0.15\,\mu\mathrm{s}$, the angularly
integrated calculation gives $f_{90}=23.3$ kHz, 233 kHz, and 1.51 MHz.  As
$\taueng$ decreases, the response shifts from tens of kilohertz toward the
megahertz range; for $\taueng=0.15\,\mu\mathrm{s}$, 25.8\% of the integrated
response energy lies between 1 and 3 MHz.  Reaching this extension requires
submicrosecond structure in the asymmetric-energy release, rather than
relativistic beaming alone; these timescales correspond to light-travel
distances of tens of metres.

We next used the GW-memory spectrum as the distributed driver of a transverse
X-mode/fast-magnetosonic response in a prescribed magnetized pair plasma.  A
global non-WKB wave equation, solved with radiative conditions excluding
incoming waves, retains both propagation directions and possible reflection.
The solution forms an outward mode, is established mainly in the inner source
region, and reaches $\rout=10^{13}\,\mathrm{cm}$.  Across the broader
0.1 kHz--30 MHz frequency range, the global transfer is consistent with the
coherent limit at the numerical resolution of the calculation.  The modeled
source-region plasma therefore does not impose an additional spectral cutoff;
the declining high-frequency response is inherited from the acceleration and
central-engine energy-release timescales that shape the GW-memory spectrum.

This work shows that a step-like GW memory can drive a frequency-resolved EM
response that escapes the modeled source region.  The central-engine timescale
and asymmetric fast-energy fraction determine whether this response is
concentrated at tens of kilohertz or extends to megahertz frequencies.  The
10 kHz--1 MHz band overlaps the frequency coverage of space-based plasma-wave
instruments \cite{Bougeret1995,Bougeret2008}, so the escaping response may in
principle be observable.  Whether it reaches the Solar System with measurable
amplitude depends on propagation through the outer magnetar environment and
intervening plasma, absolute plasma-mode normalization, instrumental
sensitivity, and the ambient plasma background.  A quantitative assessment of
the received flux, signal-to-noise ratio, event rate, and space-based
detectability is left to a follow-up study.

\begin{acknowledgments}
This work was supported by the National Science and Technology Major Project
of China (No.~2024ZD1100601), the National Key R\&D Program of China
(No.~2021YFC2203002) and the NSFC (National Natural Science Foundation of
China) No.~12473075.
\end{acknowledgments}

\bibliography{references}

\end{document}